\documentclass[sigconf]{acmart}
\usepackage{multirow}

\renewcommand\footnotetextcopyrightpermission[1]{}

\makeatletter
\def\@copyrightspace{\relax}
\makeatother

\begin{document}

\title{Hier-SPCNet: A Legal Statute Hierarchy-based Heterogeneous Network for Computing Legal Case Document Similarity}


\author{Paheli Bhattacharya}
\affiliation{\institution{IIT Kharagpur, India}}

\author{Kripabandhu Ghosh}
\affiliation{\institution{Tata Research Development and Design Centre, Pune, India}}

\author{Arindam Pal}
\affiliation{\institution{Data61, CSIRO and Cyber Security CRC \\ Sydney, NSW, Australia}}

\author{Saptarshi Ghosh}
\affiliation{\institution{IIT Kharagpur, India}}



\begin{abstract}
Computing similarity between two legal case documents is an important and challenging task in Legal IR, for which text-based and network-based measures have been proposed in literature. 
All prior network-based similarity methods considered a precedent citation network among case documents only (PCNet). However, this approach  misses an important source of legal knowledge -- the hierarchy of legal statutes that are applicable in a given legal jurisdiction (e.g., country). 
We propose to augment the PCNet with the hierarchy of legal statutes, to form a heterogeneous network Hier-SPCNet, having citation links between case documents and statutes, as well as citation and hierarchy links among the statutes. 
Experiments over a set of Indian Supreme Court case documents show that our proposed heterogeneous network enables significantly better document similarity estimation, as compared to existing approaches using PCNet. We also show that the proposed network-based method can complement text-based measures for better estimation of legal document similarity.
\end{abstract}




\maketitle
\pagestyle{plain} 

\section{Introduction}

Many countries such as India, Australia, United States and United Kingdom follow the \emph{Common Law System}, wherein there are two primary sources of law -- 
(1)~Statutes or written laws (e.g., Section 302 of Indian Penal Code which describes punishment for murder), and 
(2)~Precedents or prior cases decided by important courts (e.g., the Supreme Court, High Courts). 
In such a system, law practitioners have to look up a huge number of prior cases that match a given situation or a particular case. This calls for developing legal IR systems, such as recommendation and prior-case search systems. 

A key step for developing these legal IR systems is to {\it estimate the similarity between two legal case documents}, which 
is challenging because legal documents are long, complicated and unstructured~\cite{kumar2011similarity, kumar2013similarity, minocha2015finding, mandal2017measuring}. 
Also, there is no well defined notion of legal similarity -- two legal case documents are considered similar if legal experts judge them to be similar. In this work, we focus on the challenge of automating this similarity computation. 

Although there exists several supervised methods for {\it general} document similarity (e.g., for measuring similarity of news articles~\cite{liu2019matching}), having such supervised methods for legal document similarity is not practical. 
This is because training such supervised models need a gold standard containing thousands of similar document pairs. Since legal document similarity can be verified only by legal experts, developing such a gold standard is prohibitively expensive. 
Existing methodologies for finding similar legal documents are hence unsupervised~\cite{kumar2011similarity, kumar2013similarity, minocha2015finding, mandal2017measuring}.

The existing methods for computing legal document similarity and can be broadly classified into {\it network-based methods} that rely on citation to prior case documents~\cite{kumar2011similarity, minocha2015finding}, and 
{\it text-based methods} that rely on the textual content of the documents~\cite{mandal2017measuring}, and hybrid~\cite{kumar2013similarity}.

In this paper, we focus on network-based approaches. All existing network-based methods (including the hybrid ones~\cite{kumar2013similarity}) rely on a {\it precedent citation network} (PCNet) that capture citations from one case document to prior-case documents (see Section~\ref{sec:prec-cits}). 
However, PCNet misses an important source of legal information that is inherent in the \emph{statutes} of a particular jurisdiction (e.g., country). 
Based on what we understand from discussions with Law practitioners in India (faculty members from the Rajiv Gandhi School of Intellectual Property Law, India), statutes represent the written laws and are hence a valuable source of legal knowledge, that can be used in several tasks including estimating similarity between legal documents. 
Hence, in this work, we 
augment PCNet to construct a heterogeneous network Hier-SPCNet (Hierarchical Statute and Precedent Citation Network -- see Figure~\ref{fig:act-structure}) that encompasses the structure of the statutes as well as citation information present in them.

To estimate the similarity between legal documents, we propose to apply the graph embedding algorithm Metapath2vec~\cite{dong2017metapath2vec} on the heterogeneous Hier-SPCNet.
Our method relies on the key idea that if two documents cite a common statute/precedent or if two documents cite different statutes/precedents that are themselves structurally similar in the network, then the two documents may be discussing similar legal issues, which is a strong signal for estimating document similarity. 
We evaluate our approach on a set of $100$ document pairs comprising of case judgments from the Supreme Court of India, whose similarities have been annotated by legal experts. Results show that our proposed method achieves significant improvement over prior methods that use the PCNet alone. 

We also compare our proposed network-based method with a state-of-the-art text-based method for computing legal document similarity using document embeddings~\cite{mandal2017measuring}. We observe that the proposed network-based method can give complimentary insights compared to what is given by the text-similarity method. Combining the two is a promising way of estimating legal document similarity from multiple aspects.

To our knowledge, this is the first work that proposes a network to capture all domain information inherent in both statutes and precedents (the two main pillars of a Common Law system) and shows its utility in capturing the similarity of two legal documents. 
Also note that, though we have focused on Indian legal documents, our method can be extended to any jurisdiction that defines statutes/codes in their judicial system (e.g., France~\cite{Mazzega:2009:NFL:1568234.1568271}).

\section{Existing network-based methods for legal document similarity}
\label{sec:prec-cits}

Existing network-based similarity methods construct a {\it Precedent Citation Network} (PCNet) in which the vertices are case documents, and there is a directed edge $d_1 \rightarrow d_2$ if document $d_1$ cites another document $d_2$. 
The greyed box in Figure~\ref{fig:act-structure} shows PCNet for a small example. 
Following are the existing similarity measures applied on PCNet for finding legal document similarity:

\noindent $\bullet$ \textbf{Bibliographic Coupling~\cite{kumar2011similarity}}: It is defined as the \emph{Jaccard similarity index} between the sets of precedent citations (out-citations) from the two documents whose similarity is to be inferred.
    
\noindent $\bullet$ \textbf{Co-citation~\cite{kumar2011similarity}}: Similar to bibliographic coupling, but it is defined on the sets of {\it in-citations} from the two documents.

\noindent $\bullet$ \textbf{Dispersion~\cite{minocha2015finding}}: 
This measure measures to what extent the out-neighbours (out-citation documents) of the two documents are themselves similar, i.e., occurs in the same community/cluster. 
We use the \textit{NetworkX} implementation for this measure.\footnote{\url{https://networkx.github.io/documentation/networkx-1.9/reference/generated/networkx.algorithms.centrality.dispersion.html}}

\section{Proposed augmentation of PCNet with legal statute hierarchy}

\label{sec:hierspcnet}

We now describe how we augment PCNet using information from the legal statutes, to obtain Hier-SPCNet (Hierarchical Statute and Precedent Citation Network -- shown in Figure~\ref{fig:act-structure}), and how we use Hier-SPCNet for legal document similarity. 

\subsection{Constructing Hier-SPCNet}
\label{sub:hier}


\noindent {\bf Modeling the hierarchy of statutes:}
In most common law countries, an act has its own hierarchy. For instance, in the Indian judiciary, an act can be divided into `parts'; each `part' can be divided into `chapters'; each `chapter' can be further divided into `topics'; under a `topic' are finally `sections'/`articles'. 
An example of the Act $\rightarrow$ Part $\rightarrow$ Chapter $\rightarrow$ Topic $\rightarrow$ Section/Article hierarchy is --
{\it Constitution of India, 1950 $\rightarrow$ Part VI: The States $\rightarrow$ Chapter III: The State Legislature $\rightarrow$ Topic: Disqualification of members $\rightarrow$ Section 192: Decision on questions as to disqualification of members.}
Sometimes, for smaller acts, parts of this hierarchy may not be explicitly specified. For instance, we may have sections/articles directly under an act. An example is -- {\it Dowry Prohibition Act, 1961 $\rightarrow$ Section 3:  Penalty for giving or taking dowry.}

\begin{figure}
	\centering
	\vspace{-5mm}
	\caption{The proposed heterogeneous network Hier-SPCNet consisting of case documents and statutes. Existing methods have considered only PCNet (greyed box).}
	\label{fig:act-structure}
	\includegraphics[width=8.5cm,height=5.5cm]{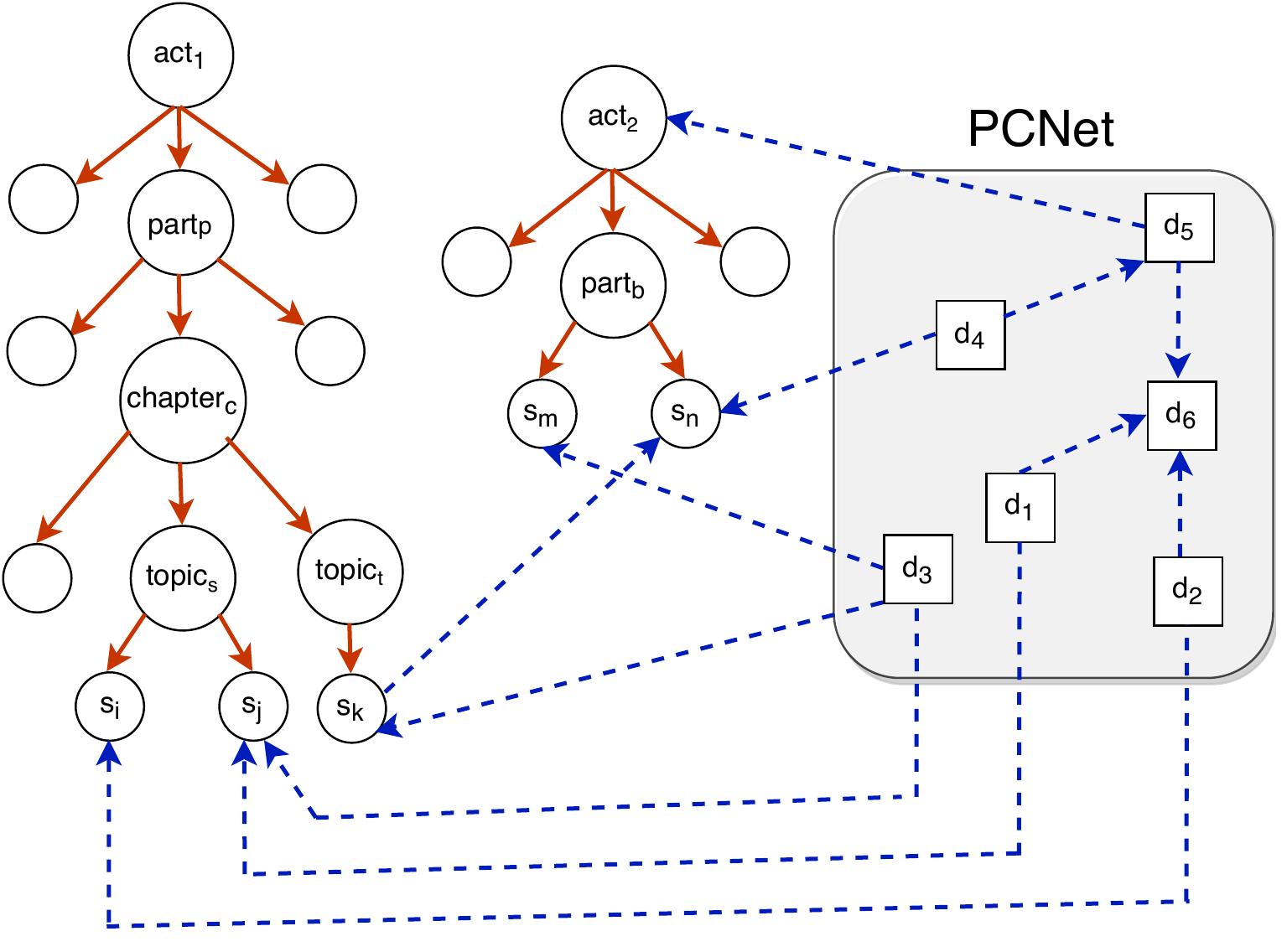}
	\label{fig:example}
	\vspace*{-10mm}
\end{figure}

For construction of Hier-SPCNet, we extract the hierarchy from the text of the statutes, and then represent each act as a hierarchical structure of nodes (act / parts / chapters / topics / sections) and hierarchy links. 
Figure~\ref{fig:act-structure} shows a pictorial representation of an act having the complete hierarchy ($act_1$) and another act having a smaller hierarchy ($act_2$).

\vspace{1mm}
\noindent {\bf Extraction of citations from text:}
Extracting statute/precedent citations from legal text is non-trivial, since the citations are written in various forms. 
We extract the citations using regular expression-based patterns, e.g.,   
the pattern $<[$section or article number$]$ of the $[$Act$]>$ is used to extract citations such as {\it `Section 47 of the Code of Criminal Procedure, 1973'}.
An internal evaluation showed that this methodology correctly extracts more than 90\% of all citations that are identified by human annotators (details omitted for brevity).

\vspace{1mm}
\noindent {\bf Hier-SPCNet:}
The network consists of \textbf{six (6) types of nodes} -- case documents, acts, parts, chapters, topics, sections (or articles).
Also there are \textbf{two types of links/edges} -- {\it hierarchy links} (orange, solid lines in Figure~\ref{fig:act-structure}) and {\it citation links} (blue, dotted lines in Figure~\ref{fig:act-structure}). 
The types of edges are described below.

\noindent $\bullet$ \textbf{Citation edges:} These edges are of three types.
\textit{(1)~ document $\rightarrow$ document}: if one document cites another document. These edges are the ones in PCNet (the grey coloured box in Figure~\ref{fig:act-structure}). Existing methods have considered only this network.
\textit{(2)~document $\rightarrow$ statute}: if a document cites a statute. For example, in Figure~\ref{fig:act-structure}, document $d_1$ cites section $s_i$ of $act_1$. 
A document can also cite an act as a whole, without referring to a particular section, e.g. document $d_5$ cites $act_2$. 
\textit{(3)~statute $\rightarrow$ statute}: if a statute cites another statute. Note that the two statutes can be part of the same or different Acts, e.g., in Figure~\ref{fig:act-structure}, statute $s_k$ of $act_1$ cites statute $s_n$ of $act_2$.

\noindent $\bullet$ \textbf{Hierarchy edges:} The hierarchy links (shown as orange, solid arrows in Fig.~\ref{fig:act-structure}) represent the hierarchy within each Act, as described in Section~\ref{sub:hier}.
These edges can be of various types, such as {\it act $\rightarrow$ part} (e.g., $part_p$ is under $act_1$ in Fig.~\ref{fig:act-structure}), {\it act $\rightarrow$ chapter}, {\it part $\rightarrow$ section} (e.g., in $act_2$, sections $s_m$ and $s_n$ are under a $part_b$), 
\textit{topic $\rightarrow$ section} (e.g., $s_i$ and $s_j$ are under $topic_s$ under $act_1$), and so on. 
Note that, as stated in Section~\ref{sub:hier}, all levels of the hierarchy may not exist uniformly in all the Acts.

\subsection{Document similarity using Hier-SPCNet}

The existing measures of bibliographic coupling, co-citation and dispersion (see Section~\ref{sec:prec-cits}) can be applied over Hier-SPCNet, similar to how they are applied over PCNet. 
However, when applied over Hier-SPCNet, these measures also include statute information, e.g. bibliographic coupling over Hier-SPCNet finds the number of common citations to prior cases as well as to statutes.

Additionally, we apply graph embedding techniques Node2Vec~\cite{node2vec} and Metapath2Vec~\cite{dong2017metapath2vec} over Hier-SPCNet. 
These embedding techniques map the nodes of the graph to a vector space, such that nodes having similar neighbourhoods in the network have similar representations (embeddings). We then compute the {\bf cosine similarity between these node embeddings} to estimate the similarity between the documents (nodes). 

\vspace{1mm}
\noindent \textbf{Node2Vec~\cite{node2vec}}: 
Given a network, Node2vec generates node embeddings (vectors) via random walks, following Breadth-First Search (BFS) or Depth-First Search (DFS). 
We apply Node2Vec on both PCNet and Hier-SPCNet.\footnote{We used the Node2vec implementation at \url{https://github.com/aditya-grover/node2vec} with embedding size of $128$ and other hyperparamaters set to default.} 
Note that Node2vec assumes a network to be homogeneous (all nodes and edges of same type). While PCNet is actually homogeneous, Hier-SPCNet is not; however, Hier-SPCNet is also considered homogeneous when applying Node2vec.

\vspace{1mm}
\noindent \textbf{Metapath2Vec~\cite{dong2017metapath2vec}}: Metapath2Vec is meant for heterogeneous networks, where nodes are of different types and the edges have different semantics. The basic working mechanism is similar to Node2Vec, but while Node2Vec uses standard BFS/DFS, Metapath2vec works on certain {\it user-defined metapaths}. 
A metapath is a path between two nodes where the edges can have different semantics. 
For Hier-SPCNet, we define {\bf 14 different metapaths} to capture situations where two documents cite the same or related statutes, whereby some signal of similarity between the documents can be inferred.\footnote{We used the implementation of Metapath2vec from  \url{https://pypi.org/project/stellargraph/}, with walk length of 5, number of random walks per root node of 2000, embedding size of 200, and other hyperparameters set to default.} 
Some of the metapaths we defined are as follows:

\noindent $\bullet$ \textbf{doc-sec-doc}: when two documents cite the same section/article. E.g., in Figure~\ref{fig:act-structure}, documents $d_1$ and $d_3$ cite the same section $s_j$.

\noindent $\bullet$  \textbf{doc-sec-topic-sec-doc}: when two documents cite different sections/articles, and the sections are under the same topic. E.g., in Fig.~\ref{fig:act-structure}, document $d_1$ cites section $s_j$ and $d_2$ cites $s_i$ and both $s_i$ and $s_j$ are under the same topic $topic_s$.

\noindent $\bullet$  \textbf{doc-sec-topic-chap-topic-sec-doc}: when two documents cite different sections, and the sections are under the same chapter. E.g., in Fig.~\ref{fig:act-structure}, $d_1$ cites section $s_j$ and $d_3$ cites $s_k$, and $s_i$ and $s_k$ are under different topics under the same $chapter_c$ of $act_1$.

\noindent $\bullet$  \textbf{doc-doc-doc}: when two documents cite a common document. This is the standard precedent citation, which is the only metapath used when applying Metapath2vec over PCNet.

\noindent Descriptions of the $10$ other metapaths are omitted for brevity.

\section{Experiments and Results}

We now describe the experiments to compare performance on various network-based methods over PCNet and Hier-SPCNet.

\subsection{Experimental setup}

\noindent {\bf Dataset used:}
We consider case documents from the Supreme Court of India, and statutes in the Indian judiciary. 
All case documents and statutes were crawled from Thomson Reuters Westlaw India (\url{http://www.westlawindia.com}). 
We used only the {\it publicly available} full texts, and did not use any proprietary information. 

The Hier-SPCNet used for the experiments, consists of $1,806$ case documents and $128$ acts (along with their hierarchies) that are cited by at least one of these documents. 
In total, there are $22,566$ nodes and $31,309$ edges in the network.
The PCNet contains the same $1,806$ case documents as nodes and 542 citation edges among the documents. 

\vspace{1mm}
\noindent {\bf Developing gold standard for document similarity:}
For evaluating methods for legal document similarity, we need a gold standard consisting of similarity scores given by legal experts for a set of document-pairs. 
To this end, two legal experts\footnote{Senior law students from the Rajiv Gandhi School of Intellectual Property Law, India} were asked to annotate the similarity of $100$ document-pairs. Each expert assigned a similarity score in the range $[0.0,1.0]$ to each document-pair, where $0.0$ indicates that the documents are entirely dissimilar, and $1.0$ indicates that the documents are very similar. 
The task of document similarity being subjective in nature, there was disagreement between the annotators for a few document-pairs, but there was reasonably good agreement for a large majority of the document-pairs. 
For a particular document-pair, we considered the mean (average) of the similarity scores given by the two annotators as the final expert similarity score.

\vspace{1mm}
\noindent {\bf Evaluation metric:}
For evaluating the performance of a particular similarity computation method, we use Pearson correlation coefficient ($\rho$) between the mean expert similarity scores and the similarity values inferred by the said method, on the 100 document-pairs. 
This metric has been used in multiple prior works on legal document similarity~\cite{kumar2011similarity, kumar2013similarity, mandal2017measuring}.

\subsection{Results: PCnet vs. Hier-SPCNet}

Table~\ref{tab:correl} shows the performance of various network-based methods on both PCNet and Hier-SPCNet. 
All the methods show statistically significant (by Student's T-Test at 95\%, $p < 0.05$) improvement when applied over Hier-SPCNet, as compared to when applied over PCNet, except for co-citation.
The value of co-citation remains the same for both networks since it depends on the common {\it in-citations}, and in-citations of documents are same in PCNet and Hier-SPCNet (since no document is cited by a statute). 
Especially, a higher value of bibliographic coupling over Hier-SPCNet highlights the fact that, for accurately estimating legal document similarity, it is important to consider citations to not only common prior-cases but also to common statutes. 

Also, there is substantial improvement for Node2Vec based similarity for Hier-SPCNet. Although Node2Vec considers the graph to be homogeneous, including the hierarchical structure of statutes over PCNet helps, since the leaf nodes, i.e., the \textit{section} nodes are structurally similar. 

The best performance is observed using Metapath2vec over Hier-SPCNet (correlation of $0.674$ with mean expert similarity score), which is able to well capture document similarity through the metapaths among the nodes. Thus, we have effectively encoded the legal knowledge inherent in the statutes though hiearchical and citation links by defining the metapath schemas.

\begin{table}[]
\caption{Pearson correlation coefficient ($\rho$) with mean expert similarity score, for similarity values inferred by various methods over the two networks. Proposed Hier-SPCNet enables statistically significantly better inference of similarity than PCNet (by Student's T-Test at 95\%). 
}
\label{tab:correl}
\begin{tabular}{|c|c|c|}
\hline
\textbf{Method}        & \textbf{$\rho$ over PCNet} & \textbf{$\rho$ over Hier-SPCNet} \\ \hline
Bibliographic Coupling & 0.279          & 0.574                \\ \hline
Co-citation            & 0.221          & 0.221                \\ \hline
Dispersion             & 0.229          & 0.287                \\ \hline
Node2Vec               & 0.448          & 0.586                \\ \hline
Metapath2Vec           & 0.215          & \textbf{0.674}                 \\ \hline
\end{tabular}
\vspace{-5mm}
\end{table}

\section{Comparing Network-based and Text-based Similarity}

Apart from network-based similarity, important signals for legal document similarity also come from the textual content of legal documents~\cite{kumar2013similarity, mandal2017measuring}. In this section, we compare the network-based and text-based methods for legal document similarity.

We consider a text-based similarity method using document embeddings (Doc2Vec), that has been shown to estimate legal document similarity better than many other methods~\cite{mandal2017measuring}. 
Following the methodology in~\cite{mandal2017measuring}, we trained a Doc2Vec model on a large set of Indian Supreme Court case judgments (which do not contain the documents in our evaluation set of $100$ document pairs). 
We then infer Doc2vec embeddings for the document pairs in our evaluation set, and compute cosine similarity between the embeddings of the documents in each pair.

\vspace{1mm}
\noindent {\bf Comparing network-based and text-based similarity:}
The text-based method (Doc2vec) achieves a correlation of $0.734$ with the mean expert similarity score (see Table~\ref{tab:text-combo}), which is slightly better than the correlation of $0.674$ achieved by the network-based method (Metapath2vec over Hier-SPCNet). The difference is {\it not} statistically significant ($p = 0.34$) by paired Student's t-test at 95\% . 
In fact, for 58 out of the 100 document-pairs, the similarity estimated by the network-based method is {\it numerically closer to the mean expert similarity score} than the similarity estimated by the text-based method, while for the other 42 document-pairs, the text-based similarity is closer to the mean expert similarity score.

We observed the document-pairs for which the text-based similarity performs better (i.e., is closer to the mean expert similarity score), and the document-pairs for which the network-based similarity performs better. We discuss below one example document-pair each of the two types.



For the document pair 1972\_31 and 1984\_115, both documents are about reservation in admission to medical colleges, and the experts have assigned a high mean similarity score of $0.85$.
The legal issues of contention are somewhat different -- while in 1972\_31 the admission criteria considers `reservation for backward classes', in 1984\_115 the criteria in argument is `domicile'. Hence, there are differences in the text, which leads to a moderate textual similarity of $0.44$. 
With respect to the statutes cited, 1984\_115 cites the `Public Employment Requirement as to Residence Act, 1957' that cites `Article 16 of the Constitution of India' which is in turn cited by 1972\_31. This follows one of our metapaths `doc-act-sec-doc'. 
Also, both the documents cite other articles that are either the same (metapath: `doc-sec-doc') or are under the same part (metapath: `doc-sec-part-sec-doc') or under the same act (metapath: `doc-sec-act-sec-doc'). 
As a result, Metapath2vec over Hier-SPCNet estimates a high similarity of $0.73$ that is much closer to the mean expert similarity score of $0.85$.

\begin{table}[]
\caption{Pearson correlation coefficient ($\rho$) with mean expert similarity score, for a text-based method~\cite{mandal2017measuring}, the proposed network-based method, and combinations of the two. None of the pairwise differences in $\rho$ is statistically significant (paired Student's T-test at 95\%).}
\label{tab:text-combo}
\vspace{-3mm}
\begin{tabular}{|c|c|}
\hline
\textbf{Method}                                                                       & \textbf{$\rho$} \\ \hline
Network-based (Metapath2vec on Hier-SPCNet) & 0.674                 \\ \hline
Text-based (Doc2Vec)                                                                  & 0.734                 \\ \hline \hline
max (text, network)                                                                   & \textbf{0.760}        \\ \hline
average (text, network)                                                               & 0.754                 \\ \hline
\end{tabular}
\vspace{-3mm}
\end{table}

Although the two methods perform comparably, an advantage of the network-based method over Doc2Vec is that it can impart some {\it explanation} to the measured similarity (elucidated by the examples above) which was duly appreciated by our legal experts.

\vspace{1mm}
\noindent {\bf Combining network-based and text-based similarity:}
The above discussion shows that the text-based and network-based methods complement each other. Hence, a combination of these two metrics seems promising. 
We tried some simple combinations using the functions \textit{average} (a pair gets the similarity value which is an average of the text-based and  network-based similarity values) and \textit{max} (a pair gets either the text-based similarity or the network-based similarity, whichever is maximum). 
The results, shown in Table~\ref{tab:text-combo}, support the idea that combining network-based and text-based measures can be beneficial, since the two methods probably capture complementary signals of legal document similarity.
Devising better methods of combination is left as future work.

\section{Conclusion}
In this work, we achieved significantly better estimation of similarity between legal documents, by developing a hierarchical network (Hier-SPCNet) comprising of the hierarchy of statutes, and then applying network embedding methods. 
To our knowledge, this is the first attempt to computationally model the legal domain knowledge inherent in the statutes, to measure legal document similarity. 
Our method would be applicable for any other jurisdiction that defines a hierarchy of statutes~\cite{Mazzega:2009:NFL:1568234.1568271}.
As a future work, we would like to develop better techniques for combining network-based and text-based similarity for legal documents.

\vspace{3mm}
\noindent {\bf Acknowledgements:} 
The authors thank the law students who helped in developing the gold standard data. The research is partially supported by SERB, Government of India, through the project `NYAYA: A Legal Assistance System for Legal Experts and the Common Man in India'. P. Bhattacharya is supported by a Fellowship from Tata Consultancy Services.

\bibliographystyle{ACM-Reference-Format}
\bibliography{references}


\begin{thebibliography}{8}


\ifx \showCODEN    \undefined \def \showCODEN     #1{\unskip}     \fi
\ifx \showDOI      \undefined \def \showDOI       #1{#1}\fi
\ifx \showISBNx    \undefined \def \showISBNx     #1{\unskip}     \fi
\ifx \showISBNxiii \undefined \def \showISBNxiii  #1{\unskip}     \fi
\ifx \showISSN     \undefined \def \showISSN      #1{\unskip}     \fi
\ifx \showLCCN     \undefined \def \showLCCN      #1{\unskip}     \fi
\ifx \shownote     \undefined \def \shownote      #1{#1}          \fi
\ifx \showarticletitle \undefined \def \showarticletitle #1{#1}   \fi
\ifx \showURL      \undefined \def \showURL       {\relax}        \fi
\providecommand\bibfield[2]{#2}
\providecommand\bibinfo[2]{#2}
\providecommand\natexlab[1]{#1}
\providecommand\showeprint[2][]{arXiv:#2}

\bibitem[\protect\citeauthoryear{Dong, Chawla, and Swami}{Dong
  et~al\mbox{.}}{2017}]%
        {dong2017metapath2vec}
\bibfield{author}{\bibinfo{person}{Yuxiao Dong}, \bibinfo{person}{Nitesh~V
  Chawla}, {and} \bibinfo{person}{Ananthram Swami}.}
  \bibinfo{year}{2017}\natexlab{}.
\newblock \showarticletitle{metapath2vec: Scalable representation learning for
  heterogeneous networks}. In \bibinfo{booktitle}{\emph{Proc. ACM SIGKDD}}.
\newblock


\bibitem[\protect\citeauthoryear{Grover and Leskovec}{Grover and
  Leskovec}{2016}]%
        {node2vec}
\bibfield{author}{\bibinfo{person}{Aditya Grover} {and} \bibinfo{person}{Jure
  Leskovec}.} \bibinfo{year}{2016}\natexlab{}.
\newblock \showarticletitle{Node2Vec: Scalable Feature Learning for Networks}.
  In \bibinfo{booktitle}{\emph{{Proc. ACM SIGKDD}}}.
\newblock


\bibitem[\protect\citeauthoryear{Kumar, Reddy, Reddy, and Singh}{Kumar
  et~al\mbox{.}}{2011}]%
        {kumar2011similarity}
\bibfield{author}{\bibinfo{person}{Sushanta Kumar}, \bibinfo{person}{P~Krishna
  Reddy}, \bibinfo{person}{V~Balakista Reddy}, {and} \bibinfo{person}{Aditya
  Singh}.} \bibinfo{year}{2011}\natexlab{}.
\newblock \showarticletitle{Similarity analysis of legal judgments}. In
  \bibinfo{booktitle}{\emph{{Proc. ACM India COMPUTE Conference}}}.
\newblock


\bibitem[\protect\citeauthoryear{Kumar, Reddy, Reddy, and Suri}{Kumar
  et~al\mbox{.}}{2013}]%
        {kumar2013similarity}
\bibfield{author}{\bibinfo{person}{Sushanta Kumar}, \bibinfo{person}{P~Krishna
  Reddy}, \bibinfo{person}{V~Balakista Reddy}, {and} \bibinfo{person}{Malti
  Suri}.} \bibinfo{year}{2013}\natexlab{}.
\newblock \showarticletitle{Similar Legal Judgements under Common Law System}.
  In \bibinfo{booktitle}{\emph{International Workshop on Databases in Networked
  Information Systems}}.
\newblock


\bibitem[\protect\citeauthoryear{Liu, Niu, Wei, Lin, He, Lai, and Xu}{Liu
  et~al\mbox{.}}{2019}]%
        {liu2019matching}
\bibfield{author}{\bibinfo{person}{Bang Liu}, \bibinfo{person}{Di Niu},
  \bibinfo{person}{Haojie Wei}, \bibinfo{person}{Jinghong Lin},
  \bibinfo{person}{Yancheng He}, \bibinfo{person}{Kunfeng Lai}, {and}
  \bibinfo{person}{Yu Xu}.} \bibinfo{year}{2019}\natexlab{}.
\newblock \showarticletitle{Matching Article Pairs with Graphical Decomposition
  and Convolutions}. In \bibinfo{booktitle}{\emph{Proc. ACL}}.
\newblock


\bibitem[\protect\citeauthoryear{Mandal, Chaki, Saha, Ghosh, Pal, and
  Ghosh}{Mandal et~al\mbox{.}}{2017}]%
        {mandal2017measuring}
\bibfield{author}{\bibinfo{person}{Arpan Mandal}, \bibinfo{person}{Raktim
  Chaki}, \bibinfo{person}{Sarbajit Saha}, \bibinfo{person}{Kripabandhu Ghosh},
  \bibinfo{person}{Arindam Pal}, {and} \bibinfo{person}{Saptarshi Ghosh}.}
  \bibinfo{year}{2017}\natexlab{}.
\newblock \showarticletitle{Measuring similarity among legal court case
  documents}. In \bibinfo{booktitle}{\emph{{Proc. ACM India COMPUTE
  Conference}}}.
\newblock


\bibitem[\protect\citeauthoryear{Mazzega, Bourcier, and Boulet}{Mazzega
  et~al\mbox{.}}{2009}]%
        {Mazzega:2009:NFL:1568234.1568271}
\bibfield{author}{\bibinfo{person}{Pierre Mazzega},
  \bibinfo{person}{Dani\`{e}le Bourcier}, {and} \bibinfo{person}{Romain
  Boulet}.} \bibinfo{year}{2009}\natexlab{}.
\newblock \showarticletitle{The Network of French Legal Codes}. In
  \bibinfo{booktitle}{\emph{Proc. Int'l Conf on Artificial Intelligence and Law
  (ICAIL)}}.
\newblock


\bibitem[\protect\citeauthoryear{Minocha, Singh, and Srivastava}{Minocha
  et~al\mbox{.}}{2015}]%
        {minocha2015finding}
\bibfield{author}{\bibinfo{person}{Akshay Minocha}, \bibinfo{person}{Navjyoti
  Singh}, {and} \bibinfo{person}{Arjit Srivastava}.}
  \bibinfo{year}{2015}\natexlab{}.
\newblock \showarticletitle{Finding Relevant Indian Judgments using Dispersion
  of Citation Network}. In \bibinfo{booktitle}{\emph{Proc. World Wide Web}}.
\newblock


\end{thebibliography}

\end{document}